\documentclass[a4paper]{article}
\RequirePackage[english]{babel}
\RequirePackage[latin1]{inputenc}
\RequirePackage[T1]{fontenc}
\RequirePackage{mathrsfs}
\RequirePackage{amsmath}
\RequirePackage{amssymb}
\RequirePackage{amsbsy}
\RequirePackage{bm}
\begin{document}
%%%%%%%%%%%%%%%%%%%%%%%%%%%%%%%%%%%%%%%%%%%%%%%%%%%%%%%%%%%%%%%%%%%%%%%%%%%%%%%%%%%%%%%%%%%%%%%%%%%
\title{\bf{The most general cosmological dynamics for ELKO Matter Fields}}
\author{Luca Fabbri\\ 
\footnotesize{INFN, Sezione di Bologna \& Dipartimento di Fisica, Universit\`{a} di Bologna}}
\date{}
%%%%%%%%%%%%%%%%%%%%%%%%%%%%%%%%%%%%%%%%%%%%%%%%%%%%%%%%%%%%%%%%%%%%%%%%%%%%%%%%%%%%%%%%%%%%%%%%%%%
\maketitle
%%%%%%%%%%%%%%%%%%%%%%%%%%%%%%%%%%%%%%%%%%%%%%%%%%%%%%%%%%%%%%%%%%%%%%%%%%%%%%%%%%%%%%%%%%%%%%%%%%%
\begin{abstract}
Not long ago, the definition of eigenspinors of charge-conjugation belonging to a special Wigner class has lead to the unexpected theoretical discovery of a form of matter with spin $\frac{1}{2}$ and mass dimension $1$, called ELKO matter field; ELKO matter fields defined in flat spacetimes have been later extended to curved and twisted spacetimes, in order to include in their dynamics the coupling to gravitational fields possessing both metric and torsional degrees of freedom: the inclusion of non-commuting spinorial covariant derivatives allows for the introduction of more general dynamical terms influencing the behaviour of ELKO matter fields. In this paper, we shall solve the theoretical problem of finding the most general dynamics for ELKO matter, and we will face the phenomenological issue concerning how the new dynamical terms may affect the behaviour of ELKO matter; we will see that new effects will arise for which the very existence of ELKO matter will be endangered, due to the fact that ELKOs will turn incompatible with the cosmological principle. Thus we have that anisotropic universes must be taken into account if ELKOs are to be considered in their most general form.
\end{abstract}
%%%%%%%%%%%%%%%%%%%%%%%%%%%%%%%%%%%%%%%%%%%%%%%%%%%%%%%%%%%%%%%%%%%%%%%%%%%%%%%%%%%%%%%%%%%%%%%%%%%
\section*{Introduction}
Quite recently, a new form of matter called ELKO has been defined; this form of matter gets its name from the acronym of the German \textit{Eigenspinoren des LadungsKonjugationsOperators}, designating spinors that are eigenstates of the charge conjugation operator: ELKO fields are spin-$\frac{1}{2}$ fermions $\lambda$ verifying the conditions $\gamma^{2}\lambda^{*}=\pm\lambda$ for self- and antiself-conjugated fields, and therefore they are a type of Majorana fields \cite{a-g/1}. This definition spells that ELKOs are topologically neutral fermions, and because a topological charge is what keeps localized the otherwise extended field, then the absence of such charges ensures that nothing protects the field from spreading, allowing them to display non-locality; as a further consequence, according to the Wigner prescription, for which fundamental fields are classified in terms of irreducible representations of the Poincar\'{e} group, ELKOs belong to non-standard Wigner classes \cite{a-g/2}.

As it turns out, ELKO fields have mass dimension $1$; this implies that they are dynamically described by second-order field equations, so that the problems regarding the Majorana mass term, usually met when dealing with first-order field equations, are in this case circumvented \cite{a-l-s/1,a-l-s/2}. On the other hand, still referring to the Wigner scheme, fundamental fields are labelled in terms of both mass and spin quantum numbers, implying that matter fields have both energy and spin density tensors, thus requiring both curvature and torsion, if ELKOs are extended to include the coupling with gravitational backgrounds \cite{b/1,b/11}.

That ELKOs have second-order differential field equations has two important consequences: first, the ELKO field equations have kinetic term with two derivatives, and secondly, their spin density tensor is differentially related to the torsion of the spacetime; these two consequences taken together imply that, within the ELKO field equations, there is the appearance of derivatives of torsion, which is itself containing derivatives of the ELKOs, and thus additional second-order derivatives of the ELKOs do arise. This circumstance tells that in the ELKO field equations, the second-order derivatives are given by the standard d'Alembertian $\nabla^{2}\lambda$ plus additional terms of the type $C^{\alpha\beta}\nabla_{\alpha}\nabla_{\beta}\lambda$ for some matrix $C^{\alpha\beta}$ that depends on torsion: the occurrence of this latter term gives rise to a situation in which, on the one hand, ELKOs may suffer acausal propagation, and on the other hand, as the field density tends to increase ELKOs may undergo singularity formation. Nevertheless, amazingly enough, all torsional fermionic back-reactions cancel exactly, so that ELKO causal propagation is preserved \cite{f/1,f/2}, while as their density increases the ELKO gravitational pull vanishes, resulting in a gravitational asymptotic freedom for which their topological non-locality is extended as to include a dynamical non-locality \cite{f}.

All these positive results constitute a successful ground upon which to develop further consequences of the ELKO theory, some of them coming from the specific features of the field and some coming from their peculiar dynamical character; the applications of ELKO models range from particle physics to cosmology, from their coupling to the Higgs field and supersymmetric extensions to candidates for dark matter and inflationary universes. All these results are discussed in references \cite{a,b/2,b-m,b-b/1,b-b/2,b-b-m-s,s/1,s/2,a-h,r-r,r-h,dr-hs,hs-dr,d-dc-hds,w-d}.

Now, because the ELKO matter field is relatively new, it still has several open problems, especially in the foundations; one of the fundamental open problems is to find the most general dynamics for the ELKO matter field: for instance, due to the presence of curvature and torsion, we have lost the commutation of the ELKO spinorial covariant derivatives, and so dynamical terms that were before reduced to the divergence of a vector, irrelevant in the action, may now be added to the lagrangian, generalizing the ELKO action. For example, one of these terms is the one given by
\begin{eqnarray}
\nonumber
&S_{\rm{additional}}
=\int D_{\mu}\stackrel{\neg}{\lambda}\sigma^{\mu\nu}D_{\nu}\lambda d\Omega\equiv\\
&\equiv\int[-Q_{\mu}(\stackrel{\neg}{\lambda}\sigma^{\mu\nu}D_{\nu}\lambda)
-\frac{1}{2}Q^{\alpha\mu\nu}(\stackrel{\neg}{\lambda}\sigma_{\mu\nu}D_{\alpha}\lambda)
-\frac{1}{4}G_{\alpha\beta\mu\nu}(\stackrel{\neg}{\lambda}\sigma^{\mu\nu}\sigma^{\alpha\beta}\lambda)]d\Omega
\end{eqnarray}
in terms of the tensors $Q^{\alpha\mu\nu}$ and $G_{\alpha\beta\mu\nu}$, irrelevant when torsion and curvature are not present; but nevertheless, in general this term does contribute to the ELKO dynamics. We may now ask if this is the only dynamical term possible, addressing the problem of finding the most general ELKO dynamics.

Since a common trend in physics is to look for generalizations of existing physical theories, this problem is already much interesting in itself; however, it also carries a consequent issue: that is how the additional dynamical terms may change the ELKO dynamics, in applications to cosmology.

In this paper, we are going to pursue a double task, first considering the fundamental problem of finding the most general dynamics of ELKO and then investigating the consequences of this generalization by studying an application to the standard model of cosmology.
%%%%%%%%%%%%%%%%%%%%%%%%%%%%%%%%%%%%%%%%%%%%%%%%%%%%%%%%%%%%%%%%%%%%%%%%%%%%%%%%%%%%%%%%%%%%%%%%%%%
%%%%%%%%%%%%%%%%%%%%%%%%%%%%%%%%%%%%%%%%%%%%%%%%%%%%%%%%%%%%%%%%%%%%%%%%%%%%%%%%%%%%%%%%%%%%%%%%%%%
\section{The most general ELKO}
In this paper, the Riemann-Cartan geometry is defined in terms of spacetime connections given by $\Gamma^{\mu}_{\alpha\sigma}$ and used to define the Riemann curvature tensor
\begin{eqnarray}
G^{\rho}_{\phantom{\rho}\eta\mu\nu}
=\partial_{\mu}\Gamma^{\rho}_{\eta\nu}-\partial_{\nu}\Gamma^{\rho}_{\eta\mu}
+\Gamma^{\rho}_{\sigma\mu}\Gamma^{\sigma}_{\eta\nu}
-\Gamma^{\rho}_{\sigma\nu}\Gamma^{\sigma}_{\eta\mu}
\end{eqnarray}
which has one independent contraction given by $G^{\rho}_{\phantom{\rho}\eta\rho\nu}=G_{\eta\nu}$ whose contraction is given by $G_{\eta\nu}g^{\eta\nu}=G$ as usual; then we define Cartan torsion tensor
\begin{eqnarray}
Q^{\rho}_{\phantom{\rho}\mu\nu}
=\Gamma^{\rho}_{\mu\nu}-\Gamma^{\rho}_{\nu\mu}
\end{eqnarray}
and the contorsion tensor is defined by
\begin{eqnarray}
K^{\rho}_{\phantom{\rho}\mu\nu}
=\frac{1}{2}\left(Q^{\rho}_{\phantom{\rho}\mu\nu}
+Q_{\mu\nu}^{\phantom{\mu\nu}\rho}+Q_{\nu\mu}^{\phantom{\nu\mu}\rho}\right)
\end{eqnarray}
with one independent contraction given by $K_{\nu\rho}^{\phantom{\rho\nu}\rho}= Q^{\rho}_{\phantom{\rho}\rho\nu}=Q_{\nu}=K_{\nu}$ as convention: by using these tensors it is possible to build the covariant derivatives, the commutator of covariant derivatives and the cyclic permutations of covariant derivatives, known as Jacobi-Bianchi identities, useful in the following. 

Our matter fields are ELKO and ELKO dual defined by
\begin{eqnarray}
&\gamma^{2}\lambda_{\pm\mp}^{*}=\eta\lambda_{\pm\mp}\ \ \ \ \ \ \ \ 
\stackrel{\neg}{\lambda}_{\pm\mp}^{*}\gamma^{2}=-\eta\stackrel{\neg}{\lambda}_{\pm\mp}
\end{eqnarray}
for which the explicit relationship between ELKO and ELKO dual is defined as
\begin{equation}
\stackrel{\neg}{\lambda}_{\mp\pm}=\pm i\lambda_{\pm\mp}^{\dagger}\gamma^{0}
\end{equation}
where $\eta=\pm1$ for self- or antiself-conjugate fields and with the label $\pm\mp$ indicating that the fields decompose into irreducible chiral projections that are eigenstates with positive/negative or negative/positive eigenvalues of the helicity operator: if we want to give to the ELKO and ELKO dual an explicit decomposition in terms of the irreducible chiral projections we may write
\begin{eqnarray}
&\stackrel{\neg}{\lambda}_{\pm\mp}=\left(\begin{array}{cc}\pm i\eta L_{\mp}^{T}\sigma^{2} \ \ \ \ \mp iL_{\mp}^{\dagger}
\end{array}\right)\ \ \ \ 
\begin{tabular}{c}
$\lambda_{\pm\mp}=\left(\begin{array}{c}L_{\pm}\\ -\eta \sigma^{2}L_{\pm}^{*}
\end{array}\right)$
\end{tabular}
\end{eqnarray}
with the label $\pm$ designating the eigenstate with positive or negative eigenvalue of the helicity operator respectively given by
\begin{eqnarray}
&\stackrel{\neg}{\lambda}_{+-}=\left(\begin{array}{cc}
-\eta q\ \ 0\ \ 0\ \ -iq^{*}
\end{array}\!\right)\ \ \ \ 
\begin{tabular}{c}
$\lambda_{+-}=\left(\begin{array}{c}
d\\
0\\
0\\ 
-i\eta d^{*}
\end{array}\!\!\right)$
\end{tabular}\\
&\stackrel{\neg}{\lambda}_{-+}=\left(\begin{array}{cc}
0\ \ -\eta p\ \ ip^{*}\ \ 0
\end{array}\!\right)\ \ \ \
\begin{tabular}{c}
$\lambda_{-+}=\left(\begin{array}{c}
0\\
b\\
i\eta b^{*}\\ 
0 
\end{array}\!\!\right)$
\end{tabular}
\end{eqnarray}
when their spin and momentum have the same direction, as discussed in references \cite{a-g/1,a-g/2}. This definition of ELKO and ELKO dual endows them with specific properties under discrete transformations, as discussed in \cite{a-l-s/1,a-l-s/2}. Being ELKO spin-$\frac{1}{2}$ spinor fields, their spinorial covariant derivatives are defined as usual, and the covariant derivatives with torsion $D_{\mu}$ are decomposable in terms of the covariant derivatives without torsion $\nabla_{\mu}$ plus torsional contributions as
\begin{eqnarray}
&\!\!\!\!D_{\mu}\!\stackrel{\neg}{\lambda}=\nabla_{\mu}\!\stackrel{\neg}{\lambda}
-\frac{1}{2}K^{\alpha\beta}_{\phantom{\alpha\beta}\mu}\stackrel{\neg}{\lambda}\sigma_{\alpha\beta}
\ \ \ \ \ \ \ \ D_{\mu}\lambda=\nabla_{\mu}\lambda
+\!\frac{1}{2}K^{\alpha\beta}_{\phantom{\alpha\beta}\mu}\sigma_{\alpha\beta}\lambda
\end{eqnarray}
in terms of the sigma matrices $\sigma_{\alpha\beta}$ along with contorsion; their commutator is
\begin{eqnarray}
&\!\![D_{\mu},D_{\nu}]\!\stackrel{\neg}{\lambda}
=\!Q^{\rho}_{\phantom{\rho}\mu\nu}D_{\rho}\!\!\stackrel{\neg}{\lambda}\!
-\frac{1}{2}G^{\kappa\iota}_{\phantom{\kappa\iota}\mu\nu}\!
\!\stackrel{\neg}{\lambda}\!\sigma_{\kappa\iota}\ \ 
[D_{\mu},D_{\nu}]\lambda\!
=\!Q^{\rho}_{\phantom{\rho}\mu\nu}D_{\rho}\lambda\!
+\!\frac{1}{2}G^{\kappa\iota}_{\phantom{\kappa\iota}\mu\nu}\sigma_{\kappa\iota}\lambda
\end{eqnarray}
in terms of the sigma matrices $\sigma_{\kappa\iota}=\frac{1}{4}[\gamma_{\kappa},\gamma_{\iota}]$ along with torsion and curvature, where the gamma matrices $\gamma_{\iota}$ belong to the Clifford algebra and from which the parity-odd gamma matrix $\gamma=i\gamma^{0} \gamma^{1} \gamma^{2} \gamma^{3}$ is defined as usual. These properties of the definition of ELKO and ELKO dual derivatives are discussed in \cite{b/1,b/11}.

At this point it is possible to employ the curvature tensors and the derivatives of the ELKO to construct invariants that will enter in the action, defining the dynamics of the model: its variation yields the field equations describing the coupling between curvature and energy density $T^{\mu\nu}$ and between contorsion and spin density $S^{\rho\mu\nu}$ as the system of gravitational field equations given by
\begin{eqnarray}
&G^{\mu\nu}-\frac{1}{2}g^{\mu\nu}G
=\frac{1}{2}T^{\mu\nu}
\label{curvature-energy}\\
&K_{\mu\alpha\beta}-K_{\mu\beta\alpha}+K_{\alpha}g_{\beta\mu}-K_{\beta}g_{\alpha\mu}
=-S_{\mu\alpha\beta}
\label{contorsion-spin}
\end{eqnarray}
where the gravitational constant has been normalized away, known as Einstein-Sciama-Kibble field equations, and the ELKO field equations. In the scheme provided by the Riemann-Cartan geometry the Jacobi-Bianchi identities above are used to see that the Einstein-Sciama-Kibble field equations are converted into conservation laws of the form
\begin{eqnarray}
&D_{\rho}S^{\rho\mu\nu}+Q_{\rho}S^{\rho\mu\nu}
+\frac{1}{2}\left(T^{\mu\nu}-T^{\nu\mu}\right)=0\\
&D_{\mu}T^{\mu\rho}+Q_{\mu}T^{\mu\rho}-T_{\mu\beta}Q^{\beta\mu\rho}
-S_{\beta\mu\kappa}G^{\mu\kappa \beta \rho}=0
\end{eqnarray}
which have to be verified, once the ELKO matter field equations are considered.

And now we are in the position to face the problem of finding the most general dynamics of ELKO matter fields, that is we have to find all dynamical terms in the ELKO matter fields that may appear in the ELKO action: to this purpose, the first thing we have to consider is that, because we do not want to get interactions of external potentials with the ELKO fields nor self-interactions between ELKO fields themselves but we want to get only dynamical terms for ELKO fields, then we cannot have anything else but derivatives of the ELKO fields; secondly, because of the fact that ELKOs have mass dimension $1$, then we have to consider two derivatives of the ELKOs. As a consequence we have that the most general dynamical term for the ELKOs is obtained by placing between the derivatives of the ELKO and ELKO dual the most general rank-$2$ matrix given by $M_{ij}$: as any matrix is writable as product of gamma matrices, then the most general rank-$2$ matrix is the most general product of two gamma matrices given by $M_{\alpha\beta} 
=(p\mathbb{I}+r\gamma)\gamma_{\alpha}\gamma_{\beta}+(q\mathbb{I}+s\gamma)\gamma_{\beta}\gamma_{\alpha}$ alone; the invariance under parity requires the absence of the parity-odd gamma matrix $\gamma$ therefore leaving the matrix $M_{\alpha\beta} 
=p\gamma_{\alpha}\gamma_{\beta}+q\gamma_{\beta}\gamma_{\alpha}$ solely. It is possible to rewrite this matrix in the following form $M_{\alpha\beta} 
=2(p-q)\sigma_{\alpha\beta}+(p+q)\mathbb{I}\eta_{\alpha\beta}$ equivalently written by renaming the parameters as $M_{\alpha\beta}=a\sigma_{\alpha\beta}+b\mathbb{I}\eta_{\alpha\beta}$ where the parameter $b$ may be normalized away with a redefinition of the fields: in the end, the most general dynamical term for ELKO matter fields is given by the action
\begin{eqnarray}
S=\int[G-D_{\alpha}\stackrel{\neg}{\lambda}(\eta^{\alpha\beta}\mathbb{I} +a\sigma^{\alpha\beta})D_{\beta}\lambda+m^{2}\stackrel{\neg}{\lambda}\lambda]d\Omega
\label{action}
\end{eqnarray}
in terms of the most general rank-$2$ matrix as the most general product of two gamma matrices depending on a single $a$ parameter. Its variation with respect to the independent fields gives the Einstein-Sciama-Kibble field equations with conserved quantities given by: the spin and energy densities
\begin{eqnarray}
\nonumber
&S_{\mu\alpha\beta}=\frac{1}{2}\left(D_{\mu}\stackrel{\neg}{\lambda}\sigma_{\alpha\beta}\lambda
-\stackrel{\neg}{\lambda}\sigma_{\alpha\beta}D_{\mu}\lambda\right)+\\
&+\frac{a}{2}\left(D^{\rho}\stackrel{\neg}{\lambda}\sigma_{\rho\mu}\sigma_{\alpha\beta}\lambda
-\stackrel{\neg}{\lambda}\sigma_{\alpha\beta}\sigma_{\mu\rho}D^{\rho}\lambda\right)
\label{spin}\\
\nonumber
&T_{\mu\nu}=
\left(D_{\mu}\stackrel{\neg}{\lambda}D_{\nu}\lambda+D_{\nu}\stackrel{\neg}{\lambda}D_{\mu}\lambda
-g_{\mu\nu}D_{\rho}\stackrel{\neg}{\lambda}D^{\rho}\lambda\right)+\\
&+a\left(D_{\nu}\stackrel{\neg}{\lambda}\sigma_{\mu\rho}D^{\rho}\lambda
+D^{\rho}\stackrel{\neg}{\lambda}\sigma_{\rho\mu}D_{\nu}\lambda
-g_{\mu\nu}D_{\rho}\stackrel{\neg}{\lambda}\sigma^{\rho\sigma}D_{\sigma}\lambda\right)
+g_{\mu\nu}m^{2}\stackrel{\neg}{\lambda}\lambda
\label{energy}
\end{eqnarray}
which is no longer symmetric because of the terms proportional to the constant $a$ of the generalized model; and the matter field equations
\begin{eqnarray}
&\left(D^{2}\lambda+K^{\mu}D_{\mu}\lambda\right)
+a\left(\sigma^{\rho\mu}D_{\rho}D_{\mu}\lambda+K_{\rho}\sigma^{\rho\mu}D_{\mu}\lambda\right)
+m^{2}\lambda=0
\label{fieldequations}
\end{eqnarray}
where again the terms proportional to the constant $a$ accounts for the generalization of the present model. When the matter field equations
\begin{eqnarray}
&D^{2}\lambda+K^{\mu}D_{\mu}\lambda+a\sigma^{\rho\mu}D_{\rho}D_{\mu}\lambda
+aK_{\rho}\sigma^{\rho\mu}D_{\mu}\lambda+m^{2}\lambda=0
\label{matter}
\end{eqnarray}
are taken into account, the geometry-matter coupling field equations given by the curvature-energy and spin-torsion field equations
\begin{eqnarray}
\nonumber
&2G_{\mu\nu}-g_{\mu\nu}G
=D_{\mu}\stackrel{\neg}{\lambda}D_{\nu}\lambda
+D_{\nu}\stackrel{\neg}{\lambda}D_{\mu}\lambda
-g_{\mu\nu}D_{\rho}\stackrel{\neg}{\lambda}D^{\rho}\lambda+\\
&+aD_{\nu}\stackrel{\neg}{\lambda}\sigma_{\mu\rho}D^{\rho}\lambda
+aD^{\rho}\stackrel{\neg}{\lambda}\sigma_{\rho\mu}D_{\nu}\lambda
-ag_{\mu\nu}D_{\rho}\stackrel{\neg}{\lambda}\sigma^{\rho\sigma}D_{\sigma}\lambda
+g_{\mu\nu}m^{2}\stackrel{\neg}{\lambda}\lambda
\label{curvature}\\
\nonumber
&2\left(K_{\mu\beta\alpha}-K_{\alpha}g_{\beta\mu}-K_{\mu\alpha\beta}+K_{\beta}g_{\alpha\mu}\right)
=D_{\mu}\stackrel{\neg}{\lambda}\sigma_{\alpha\beta}\lambda
-\stackrel{\neg}{\lambda}\sigma_{\alpha\beta}D_{\mu}\lambda+\\
&aD^{\rho}\stackrel{\neg}{\lambda}\sigma_{\rho\mu}\sigma_{\alpha\beta}\lambda
-a\stackrel{\neg}{\lambda}\sigma_{\alpha\beta}\sigma_{\mu\rho}D^{\rho}\lambda
\label{torsion}
\end{eqnarray}
become the Jacobi-Bianchi identities, therefore showing that this model is still consistently defined. As it is widely known, the energy is not symmetric and the spin is not conserved for spinor fields in general, and as it was pointed out in \cite{b/1,b/11}, compared to other spinor fields the ELKO fields have a more complex spinorial structure: thus one should expect that the energy is not symmetric and the spin not conserved \emph{a fortiori} for ELKO fields, but this occurs only in the present generalization of the ELKO model. Therefore to ensure the properties ELKO fields are expected to have, we regard this generalization as needed.

One of the first problems that now needs to be solved is the problem of the inversion of contorsion: as it is clear from equation (\ref{torsion}), the contorsion tensor is given by the spin, which is itself written in terms of derivatives containing contorsion; once in equation (\ref{torsion}) the derivatives containing contorsion are written in terms of the derivatives without contorsion plus contorsional contributions, we are left with an equation that defines contorsion only implicitly, and making contorsion explicit in this expression means to solve the contorsion inversion problem. Since the relationship that implicitly defines contorsion is linear in the contorsion itself, then this problem always admits a solution in principle, although finding it may be rather difficult in practice: in \cite{f/1,f/2,f} for the simplest theory with vanishing $a$ the problem of the inversion of contorsion has been solved completely, and here for the most general case where $a$ has general values it is possible to follow the same procedure, getting for the completely antisymmetric irreducible decomposition the expression
\begin{eqnarray}
\nonumber
&K_{\alpha\beta\mu}\varepsilon^{\alpha\beta\mu\rho}
\left[(8+(1+a)\stackrel{\neg}{\lambda}\lambda)^{2}
+(1+a)^{2}(i\stackrel{\neg}{\lambda}\gamma\lambda)^{2}\right]=\\
\nonumber
&=4(1+a)^{2}(i\stackrel{\neg}{\lambda}\gamma\lambda)
\left(\nabla_{\mu}\stackrel{\neg}{\lambda}\sigma^{\mu\rho}\lambda
-\stackrel{\neg}{\lambda}\sigma^{\mu\rho}\nabla_{\mu}\lambda\right)-\\
\nonumber
&-4(1+a)(8+(1+a)\stackrel{\neg}{\lambda}\lambda)
i\left(\nabla_{\mu}\stackrel{\neg}{\lambda}\gamma\sigma^{\mu\rho}\lambda
-\stackrel{\neg}{\lambda}\sigma^{\mu\rho}\gamma\nabla_{\mu}\lambda\right)+\\
&+3a(1+a)(i\stackrel{\neg}{\lambda}\gamma\lambda)\nabla^{\rho}(\stackrel{\neg}{\lambda}\lambda)
-3a(8+(1+a)\stackrel{\neg}{\lambda}\lambda)\nabla^{\rho}(i\stackrel{\neg}{\lambda}\gamma\lambda)
\label{dual}
\end{eqnarray}
and for the trace irreducible decomposition the expression
\begin{eqnarray}
\nonumber
&K^{\rho}\left[(8+(1+a)\stackrel{\neg}{\lambda}\lambda)^{2}
+(1+a)^{2}(i\stackrel{\neg}{\lambda}\gamma\lambda)^{2}\right]=\\
\nonumber
&=2(1+a)(8+(1+a)\stackrel{\neg}{\lambda}\lambda)
\left(\nabla_{\mu}\stackrel{\neg}{\lambda}\sigma^{\mu\rho}\lambda
-\stackrel{\neg}{\lambda}\sigma^{\mu\rho}\nabla_{\mu}\lambda\right)+\\
\nonumber
&+2(1+a)^{2}(i\stackrel{\neg}{\lambda}\gamma\lambda)
i\left(\nabla_{\mu}\stackrel{\neg}{\lambda}\gamma\sigma^{\mu\rho}\lambda
-\stackrel{\neg}{\lambda}\sigma^{\mu\rho}\gamma\nabla_{\mu}\lambda\right)+\\
&+\frac{3a}{2}(1+a)(i\stackrel{\neg}{\lambda}\gamma\lambda)
\nabla^{\rho}(i\stackrel{\neg}{\lambda}\gamma\lambda)
+\frac{3a}{2}(8+(1+a)\stackrel{\neg}{\lambda}\lambda)
\nabla^{\rho}(\stackrel{\neg}{\lambda}\lambda)
\label{trace}
\end{eqnarray}
with which the entire contorsion can be made explicit, and the contorsion inversion problem would be solved completely as well. However, obtaining a complete solution of this problem would lead us far from the central idea of the present paper, and therefore we will not try to get the inversion of contorsion, but rather, we are going to look for circumstances where (\ref{dual}) and (\ref{trace}) are the only non-vanishing irreducible decompositions of contorsion, so that contorsion would automatically be inverted, and the problem bypassed.
%%%%%%%%%%%%%%%%%%%%%%%%%%%%%%%%%%%%%%%%%%%%%%%%%%%%%%%%%%%%%%%%%%%%%%%%%%%%%%%%%%%%%%%%%%%%%%%%%%%
%%%%%%%%%%%%%%%%%%%%%%%%%%%%%%%%%%%%%%%%%%%%%%%%%%%%%%%%%%%%%%%%%%%%%%%%%%%%%%%%%%%%%%%%%%%%%%%%%%%
\section{The most general cosmological ELKO matter}
Actually, we do not need to look around very much to find a simple application to a relevant physical situation in which (\ref{dual}) and (\ref{trace}) are the only non-vanishing irreducible decomposition of contorsion, as in the standard model of cosmology one may employ the cosmological principle for the contorsion, as well as for the metric and for the matter fields, and consequently we have that considerable simplifications do occur: indeed, when both homogeneity and isotropy are implemented, we have that actually torsion is decomposable only in terms of the two vectorial irreducible parts as it was desired, while the metric is described in terms of the Friedmann-Lema\^{\i}tre-Robertson-Walker spatially flat spacetime
\begin{eqnarray}
\nonumber
&\!\!\!\!\!\!\!\!\!\!\!\!\!\!\!\!\!\!\!\!\!\!\!\!\!\!\!\!\!\!\!\!\!g_{00}=1\\
&g_{jj}=-e^{2\sigma}\ \ \ \ j=x,y,z
\end{eqnarray}
and where we can always choose the ELKO and ELKO dual as
\begin{eqnarray}
&\begin{tabular}{cccc}
$\stackrel{\neg}{\lambda}=e^{\varphi}\left(0 \ \ \ 1\ \ \ -i \ \ \ 0 \right)$
\end{tabular}\ \ \ \ 
\begin{tabular}{c}
$\lambda=e^{\varphi}\left(\begin{array}{c}0\\1\\i\\0\end{array}\right)$
\end{tabular}
\end{eqnarray}
in terms of functions of the time; notice that as an additional simplification we have that the completely antisymmetric part of contorsion happens to vanish when this particular form of the ELKO is taken. Substituting these expressions into the field equations, and then the inverted form of the torsion trace vector into all remaining field equations, we get a field equations for the curvature and for the ELKO field in terms of the $\sigma$ and $\varphi$ functions of the time alone.

At this point it is possible to solve the problem of the propagation of the ELKO field: by following the general procedure that has been outlined in references \cite{f/1,f/2}, we may now easily see that the ELKO field equation yields a characteristic equation of the form $n^{2}\equiv0$ showing that causality is always maintained; so all results obtained in \cite{f/1,f/2} are here recovered for this most general dynamical ELKO at least in the case of cosmological systems.

In the following, we will focus on the case of the ultraviolet regime, where the ELKO density field has values that are extremely high and masslessness is well approximated: in this case we have that the torsion is 
\begin{eqnarray}
&Q_{0}=3\dot{\sigma}+\left(\frac{3a}{1+a}\right)\dot{\varphi}
\end{eqnarray}
while the curvature has form
\begin{eqnarray}
\nonumber
&G^{0j}_{\phantom{0j}0j}
=\left(\frac{a}{1+a}\right)\left(\dot{\sigma}\dot{\varphi}+\ddot{\varphi}\right)\ \ \ \ j=x,y,z\\
&G^{jk}_{\phantom{jk}jk}
=-\left(\frac{a}{1+a}\right)^{2}\dot{\varphi}^{2}\ \ \ \ j,k=x,y,z
\end{eqnarray}
and with ELKO and ELKO dual field derivatives 
\begin{eqnarray}
\nonumber
&\!\!\!\!\!\!\!\!\!\!\!\!\!\!\!\!\!\!\!\!\!\!\!\!\!\!\!\!\!\!\!\!\!\!\!\!\!\!\!
D_{0}\stackrel{\neg}{\lambda}=\stackrel{\neg}{\lambda}\dot{\varphi} \ \ \ \ \ \ \ \ \ \ \ \  D_{0}\lambda=\dot{\varphi}\lambda\\
&D_{j}\stackrel{\neg}{\lambda}
=\stackrel{\neg}{\lambda}\sigma_{j0}\dot{\varphi}\left(\frac{a}{1+a}\right)\ \ \ \  D_{j}\lambda=\left(\frac{a}{1+a}\right)\dot{\varphi}\sigma_{0j}\lambda\ \ \ \ j=x,y,z
\end{eqnarray}
where the upper dot designates the partial derivative with respect to time.

And at this point it is possible to solve also the problem of singularity formation: first, we notice that, following \cite{f}, we no longer have the situation for which the ELKO fields exhibit gravitational asymptotic freedom, as seen from the fact that the curvatures do not vanish, because the energy does not vanish in the first place; then, we also remark that, following \cite{b-b/1}, we do not have the circumstance for which the ELKO fields energy and spin densities are compatible with the cosmological principle, as seen form the fact that the spatial spinorial derivatives do not vanish, because relation $D_{\mu}\lambda=-iP_{\mu}\gamma\lambda$ tells that the spatial components of the momentum do not vanish themselves: both instances are due to the fact that the parameter $a$ does not vanish, and therefore both are genuine features of this generalization. On the other hand however, if we do insist that the ELKO energy and spin densities are compatible with the cosmological principle, then the spatial spinorial derivatives must vanish, and if $a$ is allowed to have a general value, then $\dot{\varphi}=0$ and consequently $D_{\mu}\lambda\equiv0$ identically; then we have that $G_{\alpha\beta\mu\nu}\equiv0$ identically as well, and so in the ultraviolet limit we have the flatness of the torsion-metric spacetime. In this case the ELKO matter field equation and the energy-curvature field equations will be verified automatically, although the spin-torsion field equations are verified only if $\dot{\sigma}=0$, vanishing the torsion and giving the flatness of the metric spacetime. Therefore, the ELKO field will have zero energy and spin densities, and henceforth they will displaying asymptotic gravitational freedom being compatible with the cosmological principle. In this way, the results obtained in \cite{f} and \cite{b-b/1} will be recovered for this most general dynamical ELKO in the case of cosmological symmetries in the ultraviolet limit, but only in a case in which the ELKOs have no dynamics.

To draw a parallel with previous works about ELKO matter in cosmological applications we consider that in \cite{b-b/1} Boehmer and Burnett take into account ELKO matter finding that the model is compatible with the cosmological principle while here we have shown that ELKO matter in the most general model is incompatible with the cosmological principle: as it is widely known, the spin density is associated to rotational degrees of freedom that should allow only axial symmetries for the correspondingly anisotropic backgrounds, but this happens only in the present generalization of the ELKO model; to ensure the symmetries ELKO fields are expected to have we are again lead to regard this generalization as a necessary one. Correspondingly, if we want to maintain ELKOs in their utmost generality then anisotropic universes have to be considered.

Finally we recall that in \cite{a-h} Ahluwalia and Horvath discuss how the ELKO fields are local along the preferred axis of the Very Special Relativity while here we have suggested a generalization in which anisotropies arise, and therefore the ELKO locality axis may be dynamically justified within a generalization in which anisotropies are naturally present. However as the results of \cite{a-h} are not fully encompassed in this paper we are not here in the position to discuss this issue any further, and we shall defer it to future works.
%%%%%%%%%%%%%%%%%%%%%%%%%%%%%%%%%%%%%%%%%%%%%%%%%%%%%%%%%%%%%%%%%%%%%%%%%%%%%%%%%%%%%%%%%%%%%%%%%%%
%%%%%%%%%%%%%%%%%%%%%%%%%%%%%%%%%%%%%%%%%%%%%%%%%%%%%%%%%%%%%%%%%%%%%%%%%%%%%%%%%%%%%%%%%%%%%%%%%%%
\section*{Conclusion}
In this paper we have discussed the ELKO matter in the most general cosmological dynamics; we have discussed the problem of the inversion of contorsion indicating the way to solve it in general and seeing that it may be solved for in particular homogeneous-isotropic universes: we have seen that causality is always maintained, but the asymptotic gravitational freedom and the compatibility with the cosmological principle are lost if $a$ is generic. We have discussed that ELKO fields should have non-symmetric energy with non-conserved spin possessing rotational degrees of freedom that should allow only axial symmetries for anisotropic backgrounds, and we have pointed out that this occurs only if the present generalization of the ELKO model is considered. These are two fundamental reasons that lead us to regard this generalization as necessary.

The ELKO field is certainly one of the best candidates to solve many problems of the standard model of cosmology, and having ELKO fields described in terms of the most general theory is the simplest way we have to increase our explanatory power, so to embrace into a unique theory as many of the problems of the standard model of cosmology as we can. On the other hand however, anisotropic universes is what we have to pay to buy ELKOs generality.
%%%%%%%%%%%%%%%%%%%%%%%%%%%%%%%%%%%%%%%%%%%%%%%%%%%%%%%%%%%%%%%%%%%%%%%%%%%%%%%%%%%%%%%%%%%%%%%%%%%

\

%%%%%%%%%%%%%%%%%%%%%%%%%%%%%%%%%%%%%%%%%%%%%%%%%%%%%%%%%%%%%%%%%%%%%%%%%%%%%%%%%%%%%%%%%%%%%%%%%%%
\noindent \textbf{Acknowledgments.} I am deeply grateful to Prof.~Dharam Vir Ahluwalia for his valuable suggestions and kind encouragement.
%%%%%%%%%%%%%%%%%%%%%%%%%%%%%%%%%%%%%%%%%%%%%%%%%%%%%%%%%%%%%%%%%%%%%%%%%%%%%%%%%%%%%%%%%%%%%%%%%%%
%%%%%%%%%%%%%%%%%%%%%%%%%%%%%%%%%%%%%%%%%%%%%%%%%%%%%%%%%%%%%%%%%%%%%%%%%%%%%%%%%%%%%%%%%%%%%%%%%%%

%%%%%%%%%%%%%%%%%%%%%%%%%%%%%%%%%%%%%%%%%%%%%%%%%%%%%%%%%%%%%%%%%%%%%%%%%%%%%%%%%%%%%%%%%%%%%%%%%%%
\end{document}